\documentclass [twocolumn,aps,prb,showpacs]{revtex4}
\usepackage{graphicx,amssymb,amsmath,color,bm}
\pdfoutput=1

\begin{document}
\title{Thermoelectric properties of semiconductor nanowire networks}
\author{Oleksiy Roslyak$^{1,2}$}\email{oroslyak@fordham.edu}
\author{Andrei Piryatinski$^{3}$}\email{apiryat@lanl.gov} 
\affiliation{$^1$Center for Integrated Nanotechnologies (CINT) Theoretical Division, Los Alamos National Laboratory, Los Alamos, NM 87545\\
$^2$Department of Physics and Engineering Physics, Fordham University, Bronx, NY 10458\\$^3$Theoretical Division, Los Alamos National Laboratory, Los Alamos, NM 87545}

\date{\today}

\begin{abstract}
To examine thermoelectric (TE) properties of a semiconductor nanowire (NW) network, we propose a theoretical approach mapping the TE network on a two-port network. In contrast to a conventional single-port (i.e., resistor) network model, our model allows for large scale calculations showing convergence of TE figure of merit, $ZT$, with an increasing number of junctions. Using this model, numerical simulations are performed for the Bi$_2$Te$_3$ branched nanowire (BNW)  and  Cayley tree NW (CTNW) network. We find that the phonon scattering at the network junctions plays a dominant role in enhancing the network $ZT$. Specifically, disordered BNW and CTNW demonstrate an order of magnitude higher ZT enhancement compared to their ordered counterparts. Formation of preferential TE pathways in CTNW makes the network effectively behave as its BNW counterpart. We provide formalism for simulating large scale nanowire networks hinged upon experimentally measurable TE parameters of a single T-junction.
\end{abstract}
\maketitle

\section{Introduction}

Environmental concerns and demands to explore alternative energy sources have been currently motivating the search for new efficient thermoelectric (TE) materials .\cite{Bell:08,Snyder:08}  
TE efficiency is characterized by the dimensionless figure of merit, 
\begin{eqnarray}\label{ZT-def}
ZT = \frac{\sigma_{eh} S^2 T}{\kappa_{eh} + \kappa_{ph}}, 
\end{eqnarray}
which depends on the temperature, $T$, electrical conductivity associated with electrons and holes transport $\sigma_{eh}$, Seebeck coefficient, $S$, and thermal conductivity due to the electrical charge carriers, $\kappa_{eh}$, and the phonons, $\kappa_{ph}$.\cite{Mahan:97} Accordingly, enhanced performance of TE materials requires a high power factor, $\sigma_{eh} S^2$, and at the same time suppressed thermal conductivity.\cite{Snyder:08} Nanostructuring is a promising way to achieve this goal. Reduction in dimensionality of the nanomaterials can be exploited to quantize their electronic structure (e.g., resulting in density of states singularities) in order to enhance the power factor.\cite{Dresselhaus:07} The surface and interfaces facilitate phonon scattering, subsequently decreasing associated thermal conductivity.\cite{Dresselhaus:07,Mahan:94,Bera:10}

Semiconductor nanowires (NWs) are suitable candidates for reaping the TE benefits of reduced dimensionality.\cite{Davami:2011} Reduced dimensionality of NWs dramatically changes their electronic density of states which in turn was expected to improve the power factor $\sigma_{eh} S^2$. In fact, subsequent studies demonstrated that improvement in $ZT$ can rather be achieved by suppressing the phonon contribution to thermal conductivity due to boundary scattering effects, surface roughness, and lattice anisotropy.\cite{KimJungwon:2013} Specially developed NW synthesis and characterization techniques along with theoretical modeling have been applied to a broad range of semiconductor materials from groups IV, IV-VI, V-VI, II-VI, III-V to examine their TE capabilities.\cite{zhenli2012,Boukai:08,Markussen:12,Lin:03,Parker:10} Bulk doped Bi$_{x-1}$Te$_{x}$ is currently among the most efficient and commercially used TE materials with $ZT\sim1.0$ at $T\sim 300$~K.\cite{Snyder:08} However, experimental measurements on {\em large}, $\sim 50$~nm, diameter Bi$_{x-1}$Te$_{x}$ NWs report a low value of $ZT\sim 0.1$ at $T\sim400$~K.\cite{Zhou:05,Mavrokefalos:09}  Theory predicts that a reduction in diameter to $0.5$~nm should enhance Bi$_2$Te$_3$ NW performance to $ZT=14$ at $T=300$~K which rapidly decreases to $ZT\sim 1$ with the diameter increase to $\sim4$~nm.\cite{Hicks:93} The capability of growing thin Bi$_{x-1}$Te$_{x}$ NW has been demonstrated.~\cite{Trahey:2007,zhenli2012,Zhang:12}

Use of NWs in TE devices calls for scaling-up their remarkable TE properties to the macroscopic level via assembling them into functional blocks. Various architectures for TE devices utilizing parallel arrays of NWs imbedded into different matrices and array growing techniques have been proposed.\cite{Prieto:01,KimJungwon:2013} Use of NW networks with a large amount of junctions instead of parallel NW arrays can lead to an enhancement of device performance, e.g., due to additional phonon scattering at the network junctions. For example, growth of ultra-thin, $\sim 8$~nm in diameter, Bi$_2$Te$_3$ NWs with their subsequent fabrication into bulk pellets using a spark plasma sintering method has been reported.\cite{Zhang:12} The TE characterization of these pellets shows an enhanced peak $ZT\sim0.96$ at room temperature. Recently a group of researchers at Sandia reported creation of TE nanowire arrays with uniform composition along the length of Bi$_{x-1}$Sb$_{x}$ nanowires and across the spread of the NW array, which potentially can include hundreds of millions of NWs. \cite{limmer2015using}
\par
Optimization of TE devices based on NW networks requires modeling approaches capable to account for scaling up of TE properties of individual NWs to the macroscopic device level. 
The initial step in bridging these well separated length-scales is to consider the TE properties of the mesoscopic NW assemblies, which can subsequently be used as the building blocks of macroscopic devices. The large number of junctions in branched NW mesostructures makes full scale atomistic calculations not feasible, thus calling for approximate theoretical models. In one example, atomistic calculations were combined with an averaging procedure to examine the effect of {\em a single} ultra-thin ($\lesssim 2$~nm in diameter) Si NW decoration with either Si branches or chemically bound alkyl molecules.\cite{Markusen:09} This study demonstrated that for 100 branches, $ZT$ acquires an almost four fold increase as a result of enhanced phonon scattering. To address the scaling issues in modeling the NW {\em networks} in Sec.~\ref{method}, we propose a generalized approach based on mapping a NW network onto a corresponding network of  two-port TE elements and further taking advantage of electrical circuitry analysis approach.\cite{Davis:98,Choma:85,Rice:07} In Sec.~\ref{numerics}, the method is applied to examine the TE properties of model networks made of doped Bi$_2$Te$_3$ NWs. Sec.~\ref{concl} concludes our discussion. 


\begin{figure*}[t]
  \centering
  \includegraphics[width=0.65\textwidth]{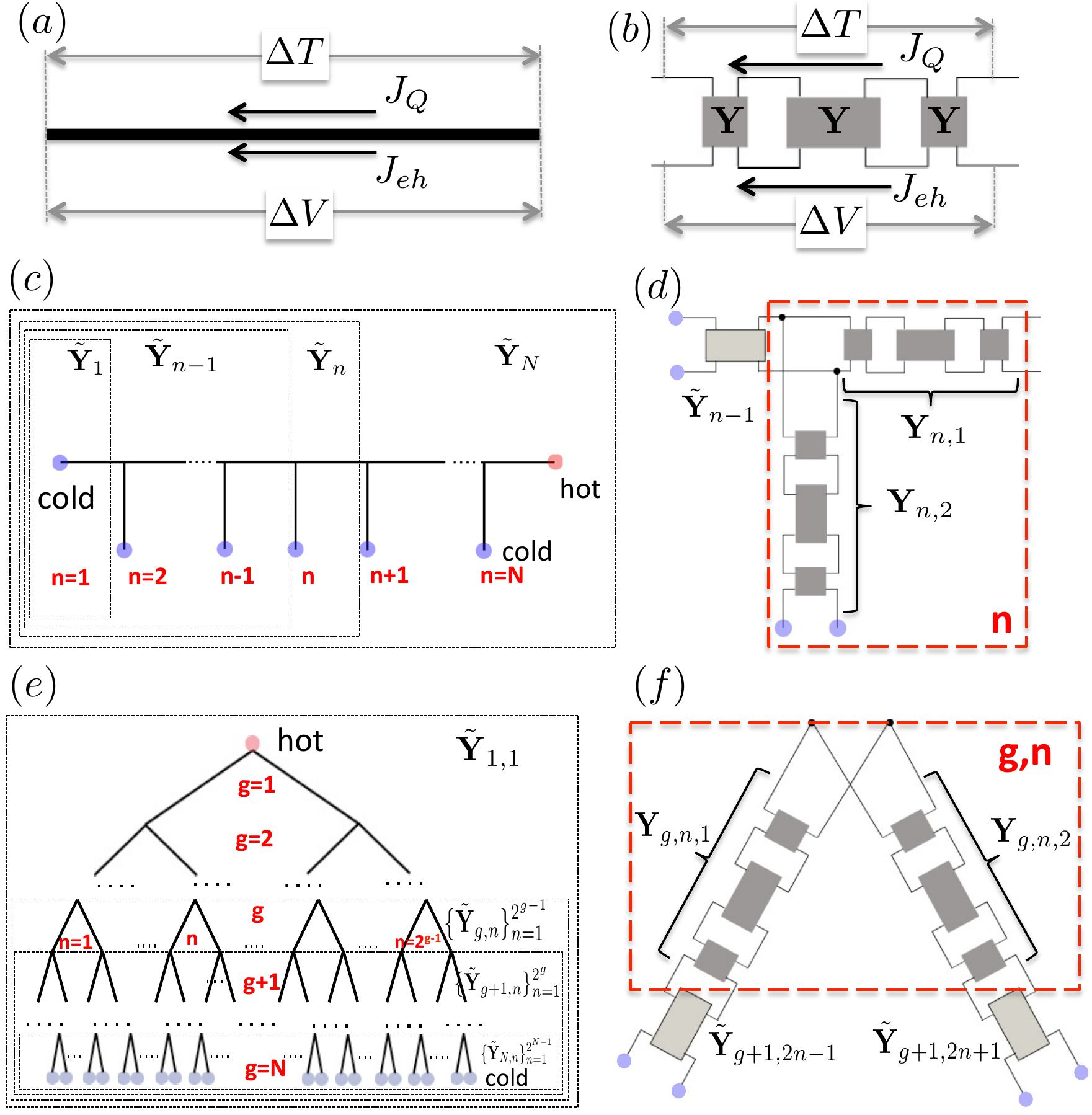}
  \caption{
(a) NW segment terminated by hot and cold junctions. (b) The segment representation as three serial connected two-port elements. (c) BNW of $N-1$ RUs and a single NW segment at site $n=1$. Dotted boxes mark BNW portions described by the admittance matrices $\tilde{\bf Y}_1,\dots,\tilde{\bf Y}_{n-1},\tilde{\bf Y}_n$. (d) Associated representation of $n$-th RU in terms of two segments from panel~(a) with the admittances ${\bf Y}_{n,1}$ and ${\bf Y}_{n,2}$. (e) CTNW of $N$ generations each containing $2^{g-1}$ RUs, where $g$ is the generation index. Dotted boxes mark the network portions each characterized by a set of admittance matrices, i.e., $\{\tilde{\bf Y}_{N,n}\}_{n=1}^{2^{N-1}}$, $\{\tilde{\bf Y}_{g+1,n}\}_{n=1}^{2^g}$, and $\{\tilde{\bf Y}_{g,n}\}_{n=1}^{2^{g-1}}$. (f) $n$-th RU from generation $g$ modeled by two NW segments from panel (a) with the admittance matrices ${\bf Y}_{g,n,1}$ and ${\bf Y}_{g,n,2}$.}
  \label{Fig-setup}
\end{figure*}

\section{NW Network Model}
\label{method}

Let us consider a TE NW network. For the temperature range of $250-400$~K used in the calculations below both electrical charge carriers and thermal transport occur in the diffusive regime leading to small local variations of electrochemical potential and temperature across the network. This allows us to partition the network into segments each of a size larger than the charge carriers and phonon mean free paths.\footnote{For  Bi$_2$Te$_3$ the mean free path for electrons is $40-60$~nm defined by the electron-phonon and impurities scattering mechanisms\cite{Mavrokefalos:09}. The phonon mean free path is typically less than that for the charge carriers.} Each segment is terminated by hot and cold junctions with other segments and/or electrical contact as shown in Fig.~\ref{Fig-setup}~(a). We assign the difference in electrochemical potential, $\Delta V = V_H-V_C$, and temperature, $\Delta T = T_H-T_C$ to each segment resulting in electric, $J_{eh}$, and thermal, $J_Q$, currents. In this case, the linear response theory yields\cite{Esfarjani:06} 
\begin{eqnarray}
\label{JYV}
\rm{\bf J} =\rm{{\bf Y}}{\rm {\bf V}} 
\end{eqnarray}
where the column vector of the TE current is 
\begin{eqnarray}
\label{J-vectr}
{\rm{\bf J}}=\left({\begin{matrix}J_{eh}\\J_Q \end{matrix}}\right) 
\end{eqnarray}
and the TE potential is
\begin{eqnarray}
\label{V-vectr}
{\rm {\bf V}}=\left({\begin{matrix}\Delta V \\ \Delta T/T\end{matrix}}\right).  
\end{eqnarray}
Taking into account the {\em two-component} nature of these vectors, we propose to use the circuit analysis approach by treating segments of a network as {\em TE two-port} elements. This allows us to identify the matrix ${\rm{\bf Y}}$ as the TE two-port admittance, and its inverse 
\begin{eqnarray}
\label{Z-def}
{\mathbf Z}={\mathbf Y}^{-1}  
\end{eqnarray}
as the TE two-port impedance.\footnote{If the electron-phonon interaction is included,\cite{Jiang:12} the proposed TE two-port model should be extended to a three-port with the extended admittance matrix.} In such a network any two segments connected in series satisfy the sum rule for the impedance matrix 
\begin{eqnarray}
\label{ser-rule}
\mathbf{Z}=\mathbf{Z}_1+\mathbf{Z}_2 
\end{eqnarray}
and the segments connected in parallel satisfy the sum rule for the admittance matrix 
\begin{eqnarray}
\label{par-rule}
\mathbf{Y}=\mathbf{Y}_1 +\mathbf{Y}_2.
\end{eqnarray}

The TE admittance matrix can be represented in terms of transport coefficients, for electrons, $L^e_i$ and holes, $L^h_i$, where $i=0,1,2$.\cite{Mermin-Book:76} By defining  $L_i=L_i^e+L_i^h$, one can write the admittance matrix as
\begin{equation}
\label{Ymtrx}
{\mathbf Y} = 
\left({
\begin{matrix}
e^2 f_{eh}(\eta) L_0& e f_{eh}(\eta) L_1\\
e f_{eh}(\eta)L_1 & f_{eh}(\eta) L_2+f_{ph}(\xi,\eta)\kappa_{ph} T
\end{matrix}
}\right),
\end{equation}
where, $e$ denotes the absolute value of the electron charge, $T= (T_H+T_C)/2$ is the averaged temperature, and $f_{eh}(\eta)$ ($f_{ph}(\eta,\xi)$) is the electron-hole (phonon) reduction function whose physical meaning is discussed below.

For a single NW, the electron/hole transport coefficients entering Eq.~(\ref{Ymtrx}) can be represented in the following general form
\begin{gather}
\label{EQ:2_2}
L^{j}_i (\mu,T) = 
\frac{2}{\hbar} \int \limits_{-\infty}^{\infty}
\mathcal{T}_{j} (E) \left({\pm E \mp \mu}\right)^i
\left({
\frac{-\partial f}{\partial E}
}\right)
d E,
\end{gather}
where $i=0,1,2$ and $j=e,h$. $\mathcal{T}_j (E)$ is the energy dependent transmission function and $f(E,\mu) = \{\exp[(E-\mu)/k_B T]+1\}^{-1}$ is the Fermi-Dirac distribution depending on the averaged chemical potential, $\mu$, and averaged thermal energy, $k_B T$. Here, the upper sign stands for the conduction band electrons ($j=e$), and the lower one for the valance band holes ($j=h$). The prefactor 2 accounts for the spin degeneracy. The transport coefficients are connected to the electrical conductivity, electron/hole contribution to thermal conductivity, and the Seebeck coefficient as \cite{Mahan-Book:00,Lin:00} 
\begin{eqnarray}
\label{eh-cond}
	\sigma_{eh} &=& e^2 L_0, \\
	\label{ph-cond}
	\kappa_{eh} &=&  \frac{L_2}{ T}-\frac{L_1^2}{ T L_0},\\
	\label{seebk}
	S &=& \frac{L_1}{e T L_0},
\end{eqnarray}
respectively. Finally, the phonon conductivity in Eq.~(\ref{Ymtrx}) can be expressed as  
\begin{gather}
\label{EQ:2_6}
\kappa_{ph} =\frac{\hbar^2}{2 \pi k_B T}\int \limits_{0}^{\infty} d \omega \omega^2\mathcal{T}_{ph}(\omega)
\frac{\exp\left(\hbar\omega/k_B T\right)}{\left[\exp\left(\hbar \omega/k_B T\right)-1\right]^2},\;\;\;\;\;\;\;\;\;\;\;
\end{gather}
where $\mathcal{T}_{ph}(\omega)$ is the phonon transmission function. 

Next, we discuss the reduction functions $f_{eh}(\eta)$ and $f_{ph}(\eta,\xi)$ entering Eq.~(\ref{Ymtrx}) that account for the junction induced drop in the electrical and thermal conductivities, respectively. Reduction in the electron/hole and phonon transmission functions associated with a single NW junction results in a drop of the electrical and thermal conductivities. To account for this effect we introduce associated  {\em single} junction transmission functions using the following equalities 
\begin{eqnarray}
\label{eta0}
\tilde{\mathcal{T}}_j(E)&=&(1-\eta)\mathcal{T}_j(E)
\\\label{xi0}
\tilde{\mathcal{T}}_{ph}(\omega)&=&(1-\xi)\mathcal{T}_{ph}(\omega), 
\end{eqnarray}
where $0\leq\eta\leq 1$ and $0\leq\xi\leq 1$ are the electron/hole and phonon reduction parameters, respectively. $\mathcal{T}_j$ ($\mathcal{T}_{ph}$) is electron/hole (phonon) transmission function of NWs forming the junction. Substitution of the junction transmission functions into Eqs.~(\ref{EQ:2_2}) and (\ref{EQ:2_6}) and comparison with Eq.~(\ref{Ymtrx}) allows one to identify single junction reduction functions as $f_{eh}(\eta)=1-\eta$ and $f_{ph}(\eta)=1-\xi$. Details on the reduction parameters determination are provided in Sec.~\ref{numerics}. Note that the junction reduction parameters and associated reduction functions are assumed to be energy independent. This approximation results in the junction independent values of the Seebeck coefficient.\footnote{According to Eq.~(\ref{seebk}), $\tilde S= \tilde L_1/(e T \tilde L_0)=f_{eh}(\eta) L_1/(e T f_{eh}(\eta) L_0)=L_1/(e T L_0)=S$.} 

Use of serial and parallel connection rules (Eqs.~(\ref{ser-rule}) and (\ref{par-rule})) for a NW network results in a complex dependance of the reduction functions entering the admittance matrix (Eq.~(\ref{Ymtrx})). In particular, the phonon reduction function becomes dependent on both electron/hole and phonon reduction parameters as indicated in Eq.~(\ref{Ymtrx}). To model NW networks, we introduce a network segment (denoted by a set of indices $s$) represented by a homogenous NW truncated by two junctions as shown in Fig.~\ref{Fig-setup}~(a). Such a segment maps on a serial connected three two-port elements shown in Fig.~\ref{Fig-setup}~(b). The central element describes the homogeneous part of the NW with $f_{eh}=f_{ph}=1$ while the right and left elements account for an identical admittance drop at the junctions. The associated reduction functions are $f_{eh}(\eta_{s})=1-\eta_{s}$ and $f_{ph}(\eta_{s},\xi_{s})=1-\xi_{s}$. Implementing the serial connection rule (Eq.~(\ref{ser-rule})), we find the following expressions for the {\em segment} reduction functions  
\begin{eqnarray}
\label{feh_n}
f_{eh}(\eta_s) &=& \frac{1}{3}-\frac{2}{9}\eta_s, 
\\
\label{fph_n}
f_{ph}(\xi_{s},\eta_{s})&=&\frac{1}{3}-\frac{2}{9}\xi_{s} 		
	-\frac{4}{9} \frac{\eta_{s}\xi_{s}(L_1^2-L_0 L_2)}{L^2_1-L_0(L_2+e^2\kappa_{ph}T)}.\;\;\;\;\;
\end{eqnarray}
Note that retaining only the linear term in Eq.~(\ref{fph_n}) yields an identical functional form for $f_{eh}$ and $f_{ph}$. This makes the electron/hole and phonon TE channels uncoupled and the network can be modeled using a conventional single-port element (i.e., resistor network) method.\cite{Markussen:12} However, in the numerical calculations (Sec.~\ref{numerics}), the cross-term is retained to account for the mixed electron/hole and phonon TE channels.  
 
Using Eqs.~(\ref{eh-cond})--(\ref{seebk}) one can easily find that the TE figure of merit (Eq.~(\ref{ZT-def})) can be expressed in terms of the TE admittance matrix as
\begin{equation}
\label{EQ:2_7}
ZT =  \frac{Y_{12}^2}{\det{\bf Y}}\;,
\end{equation}
where $Y_{12}$ denotes the off-diagonal matrix element of the admittance matrix given by Eq.~(\ref{Ymtrx}). Calculation of the figure of merit requires knowledge of the total network admittance matrix which is dependent on the network structure. Next we derive expressions allowing us to evaluate the admittance matrix for two model systems such as the branched NW (BNW) and binary Cayley tree NW (CTNW) network. 

\subsection{Branched NW }
\label{subs-BNW}

The BNW, shown in Fig.~\ref{Fig-setup}~(c), contains $\Gamma$-shaped repeat units (RU) denoted by the index $n=2,\dots,N$ and a single NW segment at the site $n=1$. All the cold contacts (blue dots) are grounded. The admittance of BNW portion containing $n-1$ RUs (dotted box) is denoted by $\tilde {\bf Y}_{n-1}$. According to panel (d), each RU contains two segments denoted by the index $m=1,2$. Each segment is characterized by the admittance matrix ${\bf Y}_{nm}={\bf Y}(\eta_{nm},\xi_{nm})$ with the reduction functions given in Eqs.~(\ref{feh_n}) and (\ref{fph_n}) where we set  $s=\{n,m\}$. By taking into account that segment $m=2$ of the $n$-th RU has a parallel connection with the rest of the network containing $n-1$ RUs and segment $m=1$ is further attached via serial connection, we use Eqs.~(\ref{Z-def})--(\ref{par-rule}) to express the admittance of the BNW portion containing $n$ RUs, $\tilde{\bf Y}_{n}$, in terms of the admittance of the $n-1$ RUs, $\tilde{\bf Y}_{n-1}$ and the admittance of the segments forming the $n$-th RU, ${\bf Y}_{n,m}$ ($m=1,2$) as
\begin{eqnarray}
\label{AlgBNW}
\tilde{\bf Y}_{n} &=& \left(\left(\tilde{\bf Y}_{n-1}+{\bf Y}_{n,2}\right)^{-1}+{\bf Y}^{-1}_{n,1}\right)^{-1}.
\end{eqnarray}
The obtained expression is exact. Starting with the initial condition $\tilde{\bf Y}_1={\bf Y}_{1,1}$ and performing $n=2,\dots N$ subsequent iterations, one can calculate the total BNW admittance $\tilde{\bf Y}_{BNW}=\tilde{\bf Y}_{N}$. Subsequently, the BNW figure of merit can be estimated by substituting ${\bf Y}=\tilde{\bf Y}_{BNW}$ into Eq.~(\ref{EQ:2_7}).

\subsection{Cayley tree NW network}
\label{sub-CTNW}

The CTNW shown in Fig.\ref{Fig-setup}~(e) contains $N$ generations each denoted by the index $g$. Each generation is formed from $\Lambda$-shaped RUs denoted by index $n=1,\dots,2^{g-1}$. For the network portion containing generations $g+1$ through $N$ (dotted box), we introduce a set of admittance matrices $\{\tilde{\bf Y}_{g+1,n}\}$ with $n=1\dots 2^g$. According to panel~(f), $n$-th RU of generation $g$ contains two segments, denoted by the index $m=1,2$. Each segment is characterized by the admittance ${\bf Y}_{gnm}={\bf Y}(\eta_{gnm},\xi_{gnm})$, where the reduction functions are given by Eqs.~(\ref{feh_n}) and (\ref{fph_n}) in which we identify $s=\{g,n,m\}$. By taking into account that both segments in each RU of generation $g$ have serial connection with the attached portion of the network (i.e., generations $g+1$ through $N$), we take advantage of Eqs.~(\ref{Z-def})--(\ref{par-rule}) to express the admittance matrix $\tilde{\bf Y}_{g,n}$ as
\begin{eqnarray}
\label{AlgCT}
\tilde{\bf Y}_{g,n} &=& \left({\bf Y}_{g,n,1}^{-1} + \tilde{\bf Y}^{-1}_{g+1,2n-1}\right)^{-1} 
\\\nonumber
&+& \left({\bf Y}_{g,n,2}^{-1} + \tilde{\bf Y}^{-1}_{g,2n+1}\right)^{-1}.
\end{eqnarray}
Similar to BNW, the obtained expression is exact and can be used for iterative calculations of the network admittance. Namely, one 
starts with the initial conditions associated with the bottom of the tree 
$\tilde{\bf Y}_{N,n}=({\bf Y}_{N,n,1}+{\bf Y}_{N,n,2})$ 
where $n=1,\dots,2^{N-1}$ and further proceeds to the top by applying Eq.~(\ref{AlgCT}) for generations $g=N-1$ through $g=1$. Eventually, the total CTNW network admittance is identified as $\tilde{\bf Y}_{CTNW}=\tilde{\bf Y}_{11}$ and the CTNW figure of merit is estimated by substituting ${\bf Y}=\tilde{\bf Y}_{CTNW}$ into Eq.~(\ref{EQ:2_7}).

\section{Numerical results and Discussion}
\label{numerics}

The numerical simulations are performed for the networks made of well parameterized $n$-doped Bi$_2$Te$_3$ NWs. Each network segment is modeled by a rectangular NW elongated in the [015] crystalline direction with the cross-section of $8\times 8$~nm$^2$. The electronic structure is calculated using a six-valance-band and six-conduction-band effective mass model accounting for the band-gap temperature dependence. Cold and hot reservoirs are modeled as infinite Bi$_2$Te$_3$ NWs. The transport coefficients  $L^e_i$, $L^h_i$ are evaluated using a constant relaxation time model. The calculations of thermal conductivity, $\kappa_{ph}$, accounts for the NW boundary specularity. Details of these calculations are given in Appendix~\ref{appx1}. 

The reduction parameters have a complex dependence on the junctions structure (e.g., lattice mismatch in the branching points). Lack of necessary microscopic parameters does not allow for their direct evaluation. Our estimate of the transmission function for Si NW junctions yields the electron/hole reduction parameter $\eta_s\sim 0.01$ and  phonon thermal conductivity reduction parameter $\xi_s\sim 0.15$. \footnote{The reduction parameters were determined according to Eqs.~(\ref{eta0}) and (\ref{xi0}) using the non-equilibrium Green function method to evaluate the transmission functions along with atomistic calculations utilizing General Lattice Utility program (GULP, http://projects.ivec.org/gulp/).} Accordingly, branching strongly suppresses the phonon assisted thermal conductivity and keeps the electrical conductivity almost intact. We adopt the same order of magnitude for the reduction parameters in Bi$_2$Te$_3$ NW networks as the {\em starting} point in our simulations.

\begin{figure}[t]
  \centering
  \includegraphics[width=0.35\textwidth]{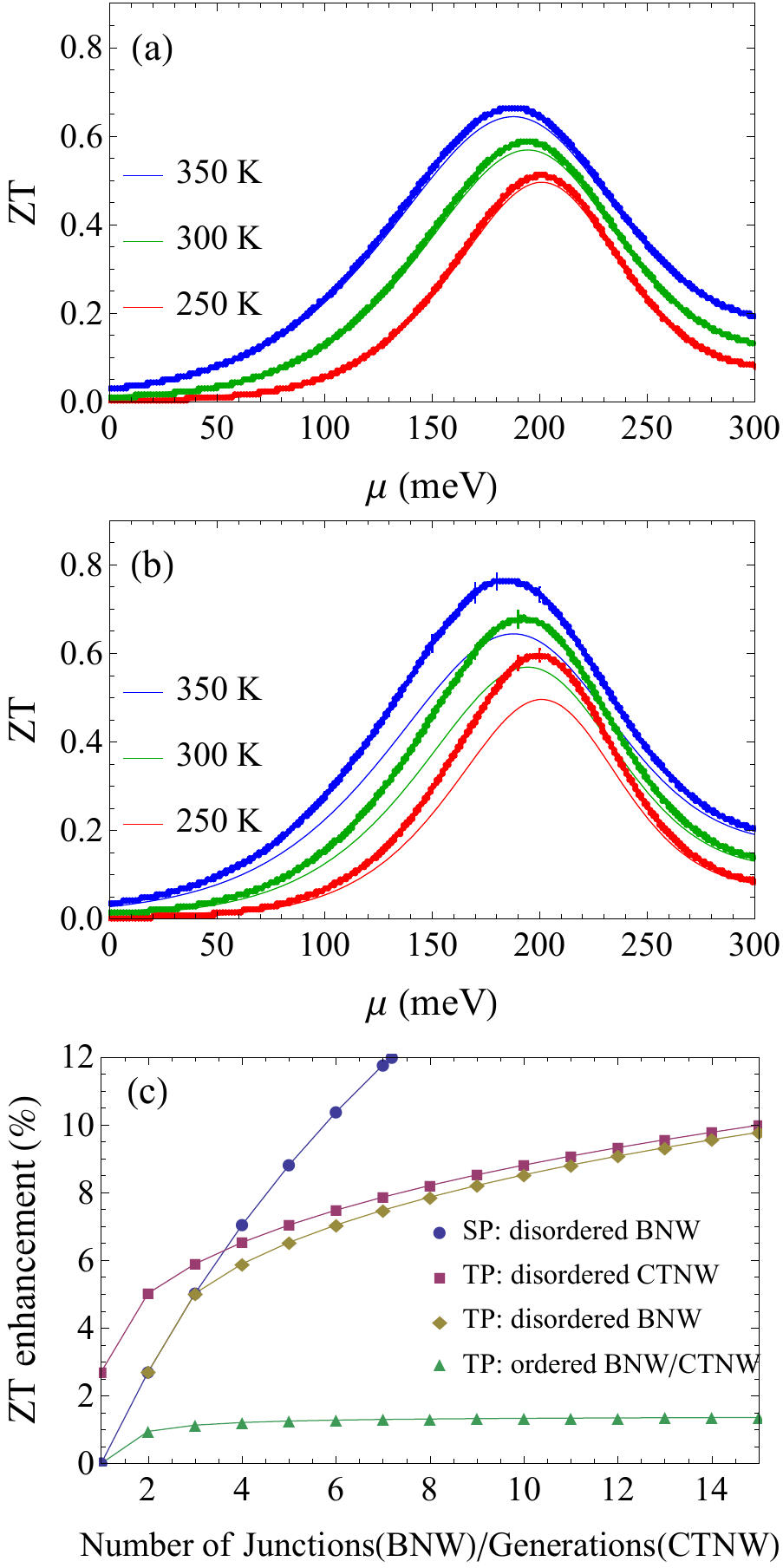}
  \caption{ZT as a function of chemical potential calculated for (a) the ordered CTNW of $N=5$ generations 
  and (b) the disordered CTNW with $N=100$ generations (thick lines). Thin lines in both panels present corresponding behavior of single NW segment. (c) ZT enhancement factor calculated at T=300~K using two-port (TP) network model for ordered/disordered BNW as a function of number of junctions (i.e., RUs$-1$) and for the  ordered/disordered CTNW as a function of number of generations. ZT vs. number of junctions in the disordered BNW calculated using single-port network model.}
  \label{Fig-ZT}
\end{figure}

Below, the following two kinds of BNW and CTNW are examined: (i) {\em Ordered} BNW/CTNW in which all junctions have identical reduction parameters $\eta_{s}=0.05$ and $\xi_{s}=0.20$. (ii) {\em Disordered} BNW/CTNW in which each junction reduction parameter is initialized according to the Gaussian distribution with the mean values $\bar\eta_{s}=0.05$ and $\bar\xi_{s}=0.20$ and $10\%$ standard deviation.

Figure~\ref{Fig-ZT}~(a) presents ZT dependence on the chemical potential, $\mu=(V_H+V_C)/2$, for the ordered CTNW (thick lines) containing $N=5$ generations. As the chemical potential and associated carriers concentration increase, the Seebeck coefficient drops while the conductivity grows resulting in a well known ZT maximum.\cite{Snyder:08} A similar plot for the disordered CTNW  of $N=100$ generations is shown in panel (b). ZT variation for single NW segment is plotted in panels (a) and (b) by thin lines for comparison. Introducing the peak enhancement factor as $$\frac{\max\limits_{\mu} \left(ZT_{\text BNW/CTNW}\right)- \max\limits_{\mu} \left(ZT_{NW}\right)}{\max\limits_{\mu}\left(ZT_{NW}\right)},$$ one finds that that branching produces about $2\%$ and $20\%$ enhancement in ZT peak values for the ordered and disordered CTNWs, respectively. Ordered and disordered BNWs with $5$ and $100$ {\em junctions} have absolutely the same ZT behavior as their CTNW counterparts. 

\begin{figure}[t]
  \centering
  \includegraphics[width=0.4\textwidth]{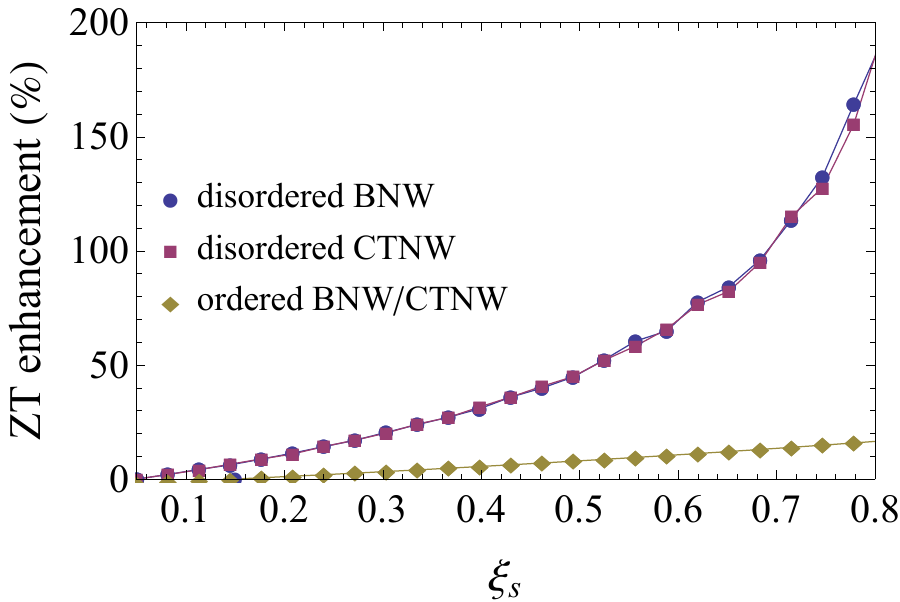}
  \caption{The enhancement factor for ordered/disordered BNW and CTNW containing 100 junctions and generations, respectively, as a function of phonon reduction parameter $\xi_s$. For disordered BNW and CTNW the mean value of $\xi_s$ is varied while the standard deviation stays $10\%$. The electron reduction parameter is fixed at $\eta_s = 0.05$ and $T=300$~K.}
 \label{Fig-ZTksi}
\end{figure}

In Fig.~\ref{Fig-ZT}~(c), we plot ZT  enhancement factor for the ordered/disordered BNW and CTNW depending on the number of {\em junctions} and {\em generations}, respectively. The ordered BNW/CTNW demonstrate identical ZT enhancement for all values of junctions/generations and saturate to $\sim 2\%$ enhancement value merely at $5$ junctions/generations. The disordered BNW/CTNW reach close ZT values at around five junctions/generation and continue monotonous growth together. Our calculations (not shown in the plot) provide saturation of the enhancement factor to $\sim 20\%$ after the number of junctions/generations exceeds $100$. 

Identical variation of ZT values for BNW and CTNW with  an increasing number of junctions and generations, respectively, suggests that within CTNW there is a preferential pathway for electrical and thermal transport. This pathway can effectively be considered as BNW whose number of junctions naturally coincides with the number of generations. By taking into account that realistic networks of NW have a more complex structure than adopted model network, one can develop a strategy of modeling realistic networks via finding the preferential pathways for TE current and mapping them on BNWs. 

Use of the NW networks in TE devices requires significant enhancement of $ZT$. Therefore, in Fig.~\ref{Fig-ZTksi} we show how the enhancement factor for 100-generation (100-junction) ordered and disordered CTNW (BNW) changes with the increase of the phonon reduction parameter. While for the disordered network this value can reach $150\%$ and higher, the ordered network shows no more than $20\%$ enhancement. The disorder effect is a reduction of effective pathways for charge carriers and phonon transport. Since, the phonon reduction parameter exceeds the carriers one by an order of magnitude, the nonlinear dependance of the phonon reduction function on this parameter amplifies the thermal current reduction and subsequently results in the drop of thermal conductivity. This  clearly demonstrates that the disordered network outperforms its ordered counterpart by about an order of magnitude in ZT enhancement and provides a good candidate for device applications.  

\begin{figure}[tt]
  \centering
  \includegraphics[width=0.3\textwidth]{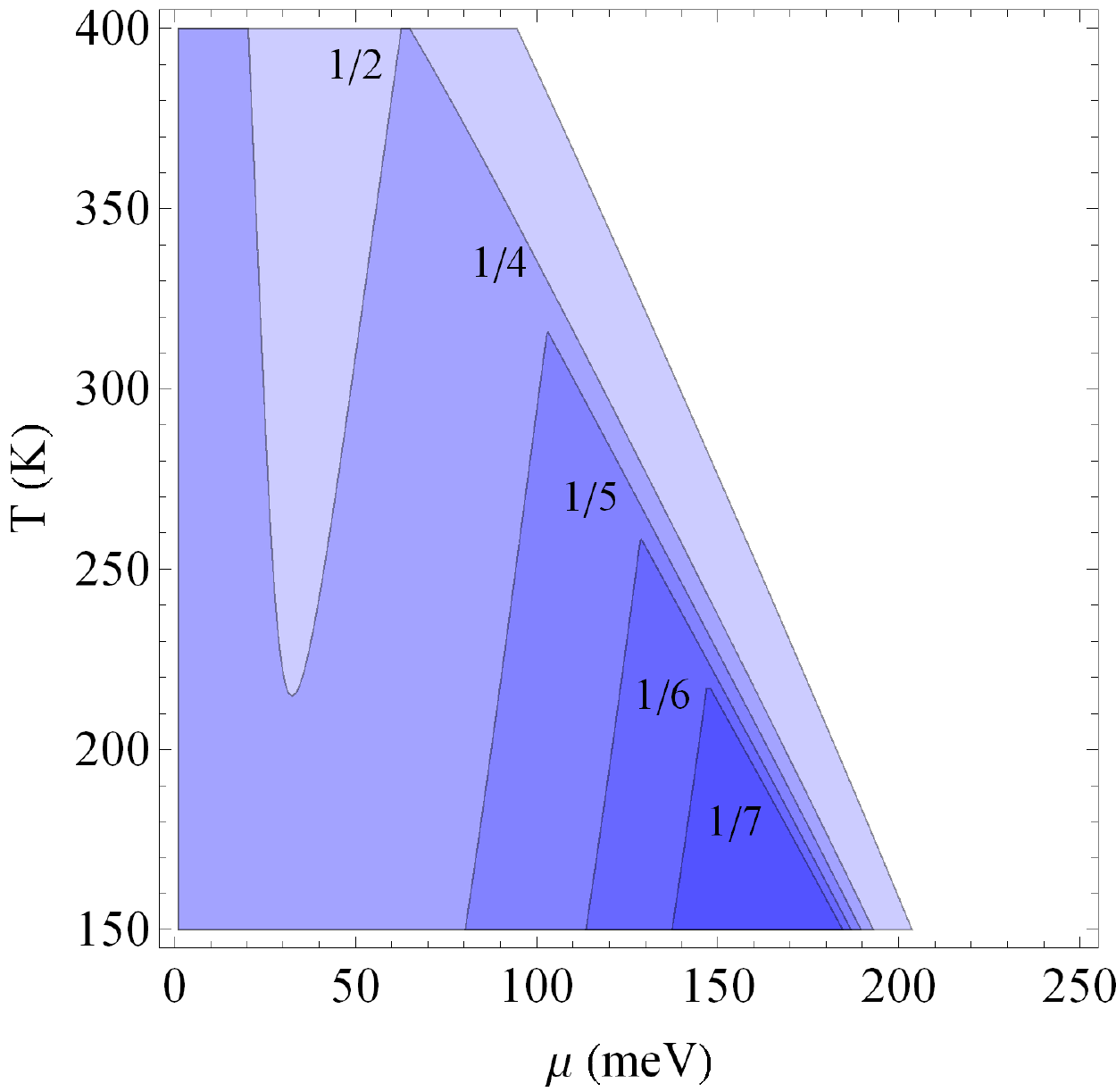}
  \caption{Temperature vs chemical potential phase digram. Lines separating areas marked with different shades of blue represent constant values of the parameter $\alpha=\min\{eL_1/e^2L_0, eL_1/\kappa_{ph}T,L_2/\kappa_{ph}T\} $ reflecting how well inequalities~(\ref{UNC1})--(\ref{UNC3}) are satisfied.}
\label{FIG:PhD}
\end{figure}

Transport properties of TE networks and devices are typically modeled by considering the electrical and thermal currents independently. This independent channel regime allows for mapping a TE network on electrical and thermal single-port (i.e., resistor) independent networks. In particular, this approach has been used to examine ZT growths with an increased number of junctions for decorated Si NWs.\cite{Markusen:09} Boundaries of the independent channel limit for our model are formulated in Appendix~\ref{appx2} and plotted in Fig.~\ref{FIG:PhD} for adopted Bi$_2$Te$_3$ NW parameters. Blue shaded area on the phase diagram covers regions where the approximation holds. The smaller value of the scaling parameter $\alpha$, the better approximation works.  

To compare our two-port model with a conventional single-port one, Fig.~\ref{Fig-ZT}~(c) presents ZT enhancement factor vs. the number of junctions in the disordered Bi$_2$Te$_3$ BNW network calculated using both models. The plot shows that both models yield the same ZT behavior for a small (about $5$) number of junctions. After that point the single-port network model shows much steeper growth than the two-port one. Our analysis reveals that in the limit of an infinite number of junctions, ZT calculated with the help of single-port network diverges. Thus we argue that the single-port network model is a good approximation for small networks whereas modeling of extended mesoscale networks should be addressed using the two-port network approach.

\begin{figure}[t]
\centering
   \includegraphics[width=0.3\textwidth]{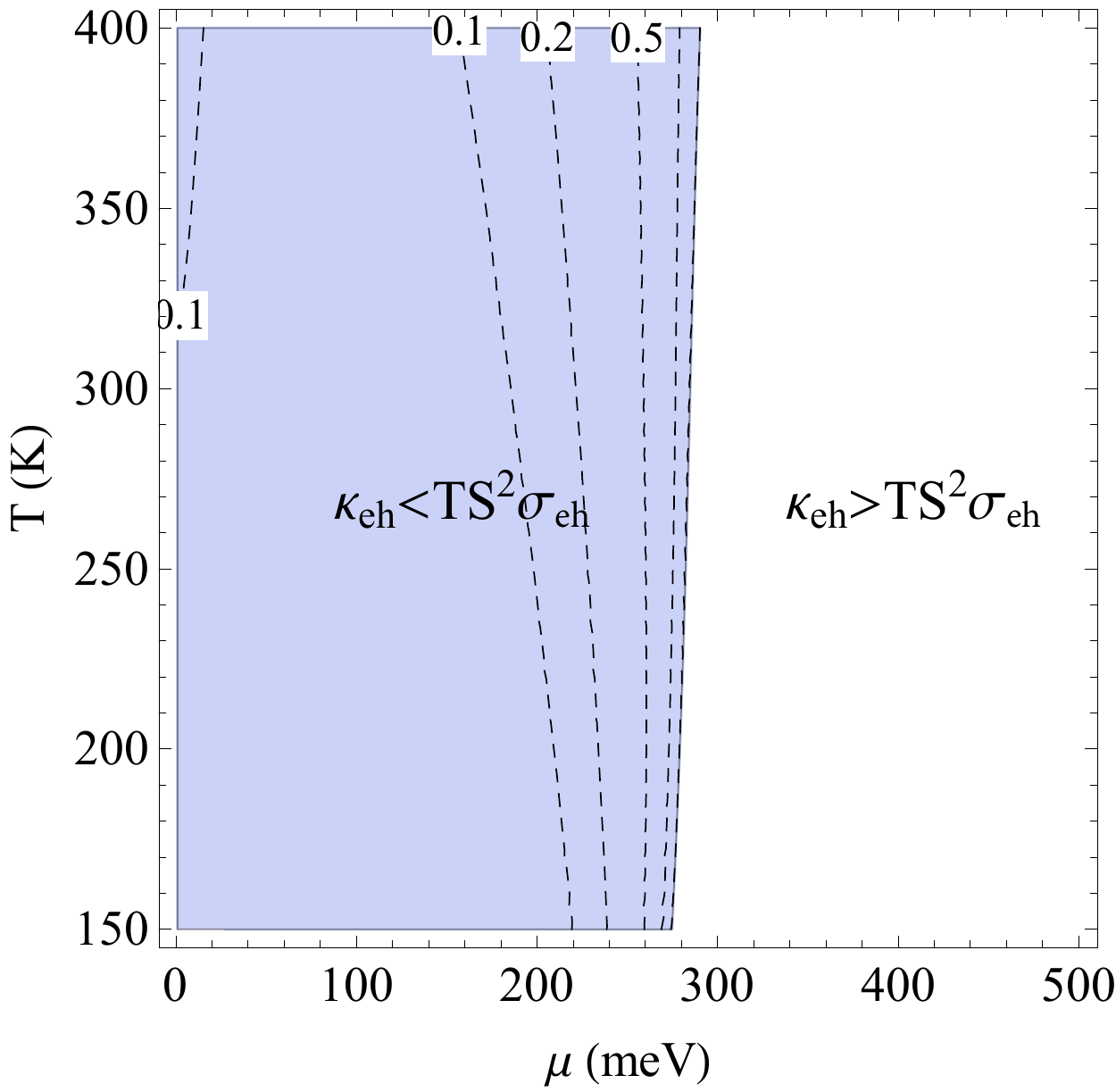}
   \caption{Temperature vs chemical potential phase digram. Dashed lines show constant values of the ratio $\kappa_{eh}/TS^2\sigma_{eh}$.}
   \label{Fig-phd}
\end{figure}

Finally, accurate modeling of TE NW networks requires precise knowledge of the junction reduction parameters. In the {\em uncoupled} TE channel regime (resistor networks) $\eta_s$ and $\xi_s$ can be directly related to the single junction impedances, $\tilde Z_{eh}$ and $\tilde Z_{ph}$ (see Appendix~\ref{appx2} for details) as
\begin{eqnarray}
\label{eta-jn}
\eta&=&1-\frac{\tilde Z_{eh}}{Z_{eh}},
\\\label{xi-jn}
\xi&=&1-\frac{\tilde Z_{ph}}{Z_{ph}},
\end{eqnarray}
where $Z_{eh}$ and $Z_{ph}$ denote homogenous NW impedances. In principle, these quantities can be determined experimentally by measuring transport coefficients entering the impedance matrices in the uncoupled channels regime. However, according to Fig.~\ref{FIG:PhD}, values of the chemical potential of $\mu \approx 200$ $\mu$eV and the temperature ranging between T=250-350K corresponding to the  maximum ZT do not correspond to the desired uncoupled regime. On the other hand, the value of Eqs.~(\ref{eta-jn}) and (\ref{xi-jn}) is that they provide a simple physical interpretation of the introduced reduction parameters. 

In a general case of two-port network, the reduction parameters can be connected to experimentally obtained values of electrical and thermal conductivities for {\em equal length} single NW and a three segment $T$-junction (Fig.~\ref{Fig-setup}~(d), $N=2$). As derived in Appendix~\ref{appx3}, the junction reduction parameters assume the simple form 
\begin{eqnarray}
\label{eta-tj}
\eta &=& \frac{2 (\sigma_{eh} - \tilde\sigma_{eh})}{2 \sigma_{eh} - \tilde\sigma_{eh}},\\
\label{xi-tj}
\xi &=& \frac{2(\kappa_{ph}-\tilde{\kappa}_{ph})}{2 \kappa_{ph} - \tilde{\kappa}_{ph}}.
\end{eqnarray}
where $\kappa_{ph}$ ($\tilde{\kappa}_{ph}$) and $\sigma_{eh}$ ($\tilde{\sigma}_{eh}$) are measured single NW (the NW T-junction) thermal and electrical conductivities, respectively. Use of the proposed expressions assumes that $\kappa_{eh}<T S^2 \sigma_{eh}$ (Eq.~(\ref{kappa-inq})). Such a regime can be reached for a specific range of the chemical potential and temperature shown as the blue region in Fig.~\ref{Fig-phd}.

\section{Conclusion} 
\label{concl}

We have developed a two-port network model to describe the TE properties of the NW networks. We argue that our model provides better approximation for mesoscale networks containing large amount of elements compared to conventional single-port (resistor) network model. Simulations based on this model are performed using parameters for Bi$_2$Te$_3$ NWs forming BNW and CTNW network. They demonstrate that the disordered networks show about an order of magnitude higher ZT enhancement compared to their ordered counterparts although it requires extended number of generations ($\sim 100$ to our estimate). Specifically for considered disordered BNW and CTNW, calculated ZT values show enhancement in the range of $20-150\%$ compared to a single NW segment. Observed similarity in ZT growth with the number of junctions and generations in BNW and CTNW, respectively. This suggests that preferential transport pathways form in CTNW effectively acting as BNWs. This provides an approach for modeling more complex network structures by mapping them on a set of BNWs. In general, use of efficient electrical circuitry simulation tools combined with our two-port TE model should  allow for direct examination of complex NW network structures. Simulations of junction reduction parameters for complex material structures is limited.  As an alternative, we provide an approach for their determination from experimental measurements on single junctions in a single-port (resistor) network limit and using a T-junction in a general case of the two-port network model.    

\acknowledgements O.R. acknowledges the support provided by the Center for Integrated Nanotechnologies (CINT), a U.S. Department of Energy, Office of Basic Energy Sciences (OBES) user facility. AP acknowledges the support provided by Los Alamos National Laboratory Directed Research and Development (LDRD) Funds. We thank Jennifer Hollingsworth for stimulating discussions.

\appendix
\section{Microscopic expressions of TE admittance matrix elements}
\label{appx1}

In this appendix, we present microscopic expressions for the transport coefficients entering TE admittance matrix given by Eq.~(\ref{Ymtrx}). These coefficients describe the charge carrier contributions to the electrical and thermal conductivity and the Seebeck coefficient. Representation for the phonon-assisted thermal conductivity entering the TE admittance matrix is provided as well. Finally, we discuss parameterization of Bi$_2$Te$_3$ NW.

In the constant relaxation time approximation (valid at $T>200 K$), the Drude conductivity model yields\cite{Lin:00,Benjenari:08}
\begin{eqnarray}
\label{EQ:7_1}
L^j_0 &=& C_j \sum_{\alpha,\beta} \mathcal{F}_{-1/2}\mu^j_{\alpha\beta},\\
\label{EQ:7_2}
L^j_1 &=&  C_j k_B T\sum_{\alpha,\beta}\left\{{3 \mathcal{F}_{1/2}\mu^j_{\alpha\beta} 
\mp \mu^j_{\alpha\beta} \mathcal{F}_{-1/2}}\mu^j_{\alpha\beta}\right\},\;\;\\
\label{EQ:7_3}
L^j_2 &=&  C_j(k_B T)^2\sum_{\alpha,\beta}
\{ 5\mathcal{F}_{3/2}\mu^j_{\alpha\beta} \mp 6\mu^j_{\alpha\beta}\mathcal{F}_{1/2}\mu^j_{\alpha\beta}
 \\ \nonumber
 &+&(\mu^j_{\alpha\beta})^2\mathcal{F}_{-1/2}\mu^j_{\alpha\beta}\},
\end{eqnarray}
where the integer indices $\alpha,\beta$ stay for the quantum numbers associated with the size quantization across the NW. 
The prefactor $C_j= ({2 e \tilde{\mu}_j m^*_{j,y}/\pi A \hbar})\sqrt{2 m^*_{j,y} k_B T}$ contains the NW cross-section, $A$, carriers effective mass along the NW axis (identified as $y$-axis) $m^*_{j,y}$, and  the carriers mobility $\tilde{\mu}_j$. The latter quantity is evaluated according to the Matthiesson's rule, $1/\tilde{\mu}_j=(1+ (\Lambda/\sqrt{A/\pi}) (1-p)/(1+p))/\tilde{\mu}^{b}_j$, using as  parameters bulk carrier mobility, $\tilde\mu^{b}_j$, mean free path $\Lambda$, and surface specularity $p$.\cite{Mavrokefalos:09} In Eqs.~(\ref{EQ:7_1})--(\ref{EQ:7_3}), 
$$\mathcal{F}_m(\mu)= \int \limits_{0}^{\infty} \frac{x^m dx}{\exp(x\mp\mu)+1}$$
 is the Fermi-Dirac fractional integral whose argument is the reduced chemical potential, $\mu^j_{\alpha\beta} = (\mu-E^j_{\alpha\beta}(0))/k_B T$, depending on the carriers energy at the band edge of associated dispersion curve, $E^j_{\alpha,\beta}(0)=E^j_{\alpha,\beta}(k=0)$, characterized by the carriers wave vector, $k$, along the NW direction. Finally, summation over all degenerate electron/hole pockets must be carried out in Eqs.~(\ref{EQ:7_1})--(\ref{EQ:7_3}).

To model the electronic structure of Bi$_2$Te$_3$ NW, we adopted the effective mass $\bf k\cdot p$ approach. Specifically, rectangular Bi$_2$Te$_3$ NW grown along  $[015]$ crystallographic direction is considered.\cite{Zhang:12} As mentioned above, the NW axis is aligned along the $y$-direction so that the $(x,z)$-plane contains the NW cross-section. The sides of the rectangle are set to be identical, $a_z=a_x = 8$~nm. According to this model the electron/hole energy at the band edge is
\begin{equation}
E^j_{\alpha,\beta}= \pm E_g(T)/2 \pm \frac{\hbar^2 \pi^2}{2} \left({\frac{\alpha^2}{m^*_{j,x} a^2_x}+\frac{\beta^2}{m^*_{j,z} a^2_z}}\right).
\end{equation}
Here, zero energy is set at the middle of the bulk band gap. The temperature dependent values of the effective masses $m^*_{j,n}$ with $j=e,h$ and $n=x,y,z$ are adopted from the literature.\cite{Benjenari:08} The band gap temperature dependence is $E_g(T) = 150.0 - 0.0947 T^2/ (T + 122.5)$~meV.\cite{Nag:80} Finally, the bulk mobilities are set to $\tilde{\mu}^{b}_e = 652.8$~cm$^2$/Vs and $\tilde{\mu}_h^b = 345.9$~cm$^2$/Vs.\cite{Benjenari:08,Goltsman:72,Goldsmith:10} and scale with the temperature as $\tilde{\mu}_e^{b}\sim T^{-1.7}$ and $\tilde{\mu}_h^b\sim T^{-2.0}$.\cite{Fleurial:88}  The adopted specularity parameter is $p=0.95$ and the mean free path of the charge carriers is set to $\Lambda \approx 40$~nm.\cite{Mavrokefalos:09}

The boundary roughness described by the specularity parameter $p$ is crucial for the phonon transport in Bi$_2$Te$_3$ NW.\cite{Mavrokefalos:09} To account for the specularity effect, solution of the Boltzmann equation is complimented  with the corresponding boundary condition. This along with the Drude conductivity in place of $\mathcal{T}_{ph}(\omega)$ results in \cite{Zou:01}
\begin{eqnarray}
\label{EQ:3_1}
\kappa_{ph} &=&
\left({\frac{k_B}{\hbar}}\right)^3
\frac{k_B T^3 \Lambda_{ph}}{2 \pi^2 v^2}
\left({1-G(p)}\right)
\\\nonumber&\times&
\int \limits_0^{\Theta_D/T} 
\frac{\epsilon^4 \exp(\epsilon) d\epsilon}{\left({\exp(\epsilon)-1}\right)^2},
\end{eqnarray}
where the specularity function is 
\begin{eqnarray}
G(p) &=& \left({1-p}\right)^2
\sum \limits_j j p^{j-1}
 \int \limits_0^1 dx\sqrt{1-x^2}
\\\nonumber &\times&
\int_0^{\pi/2}d\theta
\exp\left({\frac{-2 j x\sqrt{A/\pi}}{\Lambda_{ph} \sin \theta}}\right) \cos^2 \theta \sin \theta  .
\end{eqnarray}
Here $v$ is the average value of the acoustic phonon group velocity weighted by the phonon population factor. $\Lambda_{ph}$ is the phonon mean free path, $\epsilon = \hbar \omega/ k_B T$ is normalized phonon energy, and $\Theta_D$ is the Debye temperature. For the narrow wires ($\Lambda_{ph} \leq \sqrt{A}$ where   $A$ is the NW cross-section), the group velocity and the mean free path become functions of $\epsilon$ due to quantization of the phonon energy. However here, we consider the situation in which $\sqrt{A} \gg \Lambda_{ph}$ so that the size quantization of phonons becomes negligible and the specularity effect dominants.

In the numerical simulations of Bi$_2$Te$_3$ NWs the following parameters are used: The transverse, longitudinal, and twisting branches of the acoustic phonons are characterized by their group velocities $v=8.47\times 10^{5}$, $5.34\times 10^{5}$, and $3.95\times 10^{5}$~cm/s, respectively. The phonon mean free path is set to $\Lambda_{ph} \approx 2.5$~nm,\cite{Mavrokefalos:09} and the Debye temperature is chosen so that $\kappa_{ph} \approx 450/T$ for $T > 200 K$, and $p=0.95$.\cite{Nag:80} .


\section {Single-port network limit}
\label{appx2}

We restrict our model to the regime of uncoupled electrical and phonon channels facilitating single-port (resistor) network approximation. This approximation is valued under the conditions 
\begin{eqnarray}
\label{UNC1}
e L_1 &\ll& e^2 L_0 ,\\
\label{UNC2}
e L_1 &\ll& \kappa_{ph} T,\\
\label{UNC3}
L_2 &\ll& \kappa_{ph} T,
\end{eqnarray}
simplifying the admittance matrix (Eq.~(\ref{Ymtrx})) to the following diagonal form
\begin{equation}
\label{Y-uncpl}
{\bf Y} = 
\left({
\begin{matrix}
e^2 f_{eh}(\eta) L_0& 0\\
0 & f_{ph}(\xi)\kappa_{ph} T
\end{matrix}
}\right).
\end{equation}
Accordingly, the electrical and phonon impedances of a NW (i.e., $f_{eh}(\eta)=f_{ph}(\xi)=1$) can be defined as $Z_{eh}=({\bf Y}^{-1})_{11}=(e^2 L_0)^{-1}$ and $Z_{ph}=({\bf Y}^{-1})_{22}=(\kappa_{ph} T)^{-1}$, respectively. For a single junction, the corresponding impedances become defined as $\tilde Z_{eh}=(\tilde{\bf Y}^{-1})_{11}=(e^2 f_{eh}(\eta) L_0)^{-1}$ and $\tilde Z_{ph}=(\tilde{\bf Y}^{-1})_{22}=(f_{ph}(\xi)\kappa_{ph} T)^{-1}$. Substitution of the reduction functions in the form $f_{eh}(\eta_{s})=1-\eta_{s}$ and $f_{ph}(\eta_{s},\xi_{s})=1-\xi_{s}$ into these expressions immediately results in  Eqs.~(\ref{eta-jn}) and (\ref{xi-jn}).
 
\section{Relationship between reduction parameters and electrical and thermal conductivities for NW T-junction}
\label{appx3}

In this Appendix, we examine a T-shaped junction made of three {\em identical} NWs. For such a junction, we denote the electrical conductivity $\tilde{\sigma}_{eh}$, thermal electron/hole-assisted conductivity $\tilde{\kappa}_{eh}$, thermal phonon-assisted conductivity $\tilde{\kappa}_{ph}$, and the Seebeck  coefficients $\tilde{S}$. By using general relations (Eqs.~(\ref{eh-cond})--(\ref{seebk})) between transport coefficients and the electrical conductivity, we represent the T-junction admittance matrix (Eq.~(\ref{Ymtrx})) as
\begin{equation}
\label{EQ:MATRIX1}
\tilde{\bf Y}_{2}= 
\left({
\begin{matrix}
\tilde\sigma_{eh}/e^2 & T \tilde S \tilde\sigma_{eh}/e \\
T \tilde S \tilde\sigma_{eh}/e & T^2\tilde{S}^2 \tilde{\sigma}_{eh} +T\tilde{\kappa}_{eh}+T\tilde{\kappa}_{ph}
\end{matrix}
}\right).
\end{equation}
Note that the effect of the junction is already included into the transport coefficients ($\tilde L_i$, $i=0,1,2$) and no reduction functions are needed in Eq.~(\ref{EQ:MATRIX1}). 

On the other hand, we introduce similar parameters for each NW forming the T-junction as $3\sigma_{eh} = e^2 L_0$, $S= L_1/(e T L_0)$, $3\kappa_{eh} =   L_2/e^2 T-L_1^2/(e^2 TL_0)$, and $3\kappa_{ph}$. Here, the factor of 3 comes from the fact that each NW segment has three times larger conductivity than the whole NW. Further applying Eq.~(\ref{AlgBNW}) for the T-junction with $N=2$ along with the reduction functions in the from of Eqs.~(\ref{feh_n}) and (\ref{fph_n}), one obtains

\begin{widetext}
\begin{equation}
\label{EQ:MATRIX2}
\tilde{\bf Y}_{2}= 
\left({
\begin{matrix}
\frac{2 \sigma_{eh} (\eta -1)}{e^2 (\eta -2)} & \frac{2 S T \sigma_{eh} (\eta -1)}{e (\eta - 2)} \\
\frac{2 S T \sigma_{eh} (\eta -1)}{e (\eta - 2)} & {\frac{2T^2 S^2 \sigma_{eh} (\eta - 1)}{\eta - 2} + \frac{2T(\kappa_{ph}+ \kappa_{eh})\left({(\xi - 1) \kappa_{ph} + (\eta -1 )\kappa_{eh}}\right)}{(\xi -2) \kappa_{ph} + (\eta - 2) \kappa_{eh}}}
\end{matrix}
}\right).
\end{equation}
\end{widetext}

Comparing the admittance matrix elements in Eq.(\ref{EQ:MATRIX1}) with  Eq.(\ref{EQ:MATRIX2}), one arrives at the following expression for the reduction parameters in terms of the introduced electrical and thermal conductivities 
\begin{eqnarray}
\label{eta-final}
\eta &=& \frac{2 (\sigma_{eh} - \tilde\sigma_{eh})}{2 \sigma_{eh} - \tilde\sigma_{eh}},\\
\label{xi-final}
\xi &=& 2 + \frac{2 \kappa_{eh} \sigma_{eh}}{ 2 \kappa_{ph} \sigma_{eh} - \kappa_{ph} \tilde\sigma_{eh}}
\\\nonumber&+& 
\frac{2 \left({\kappa_{ph}+\kappa_{eh}}\right)^2}{\kappa_{ph}\left({\tilde{\kappa}_{ph}+\tilde{\kappa}_{eh} - 2(\kappa_{ph}+\kappa_{eh})}\right)}.
\end{eqnarray}
Deriving this expression, we used the fact that the junction does not change the Seebeck coefficient, i.e., $S = \tilde{S}$. As a result Eqs.~(\ref{eta-final}) and (\ref{xi-final}) become independent of $S$. To uncouple the electron-assisted and phonon-assisted effects on thermal conductivity, we impose a mild constraint 
\begin{gather}
\label{kappa-inq}
\kappa_{eh} \ll S^2 T \sigma_{eh}. 
\end{gather}
This conduction eliminates $\tilde\kappa_{eh}$ and $\kappa_{eh}$ in the matrix element $(\tilde{\bf Y}_{2})_{22}$ of Eq.~(\ref{EQ:MATRIX1}) and (\ref{EQ:MATRIX2}), respectively. Subsequently, one can set $\tilde\kappa_{eh}=\kappa_{eh}=0$ in Eq.~(\ref{xi-final}) resulting in
\begin{gather}
\label{xi-simp}
\xi = \frac{2(\kappa_{ph}-\tilde{\kappa}_{ph})}{2 \kappa_{ph} - \tilde{\kappa}_{ph}}.
\end{gather}
Obtained Eqs.~(\ref{eta-final}) and (\ref{xi-simp}) are used in the main text as Eqs~(\ref{eta-tj}) and (\ref{xi-tj}), respectively.


\bibliographystyle{prsty}

\end{document}